# Reliable Spread Spectrum Communication Systems by Using Low-Density Parity-Check Codes

Hadi Khodaei Jooshin
Department of Electrical & Computer Engineering
University of Tabriz
Tabriz, Iran
Hadikhodaei.j@gmail.com

Mahdi Nangir
Department of Electrical & Computer Engineering
University of Tabriz
Tabriz, Iran
nangir@tabrizu.ac.ir

*Abstract*—A joint scheme for the channel coding and spectrum spreading communication system is proposed in this paper. The Bit Error Rate (BER) performance of the joint proposed scheme is evaluated and compared with the case of channel coding scheme without spectrum spreading. We employ a single Low-Density Parity Check (LDPC) code for the both channel coding and direct sequence spectrum spreading. We show that the BER performance is significantly improved while the complexity does not increase too. The philosophy behind this achievement is that the spectrum spreading techniques provide the ability of channel noise protection for the communication systems by themselves.

*Keywords—LDPC Codes; Direct Sequence; Spread Spectrum; SPA; Channel Coding; BER.*

I. INTRODUCTION

By growing the technology and increasing the use of wireless communication systems, data encounter some problems during transmission process in the communication systems and channels such as jamming, interception, high-power noise, etc. [1]. This causes to increase the demand for a high quality and strong telecommunication system. Now, the question is, what do we have to do to overcome on these destructive factors and their bad effects?

The procedures that is going to be perused in this paper is combining LDPC codes and Direct Sequence Spread Spectrum (DSSS) systems [2] together for generating the spreading sequence with the capability of channel error correction. We utilize the parity-check matrix $H$ of an LDPC code for producing spreading codes. Basically, the LDPC codes are one of the efficient channel error correction codes that is being used in communication systems abundantly [3]. These codes are linear block codes that aim to compensate the corruption caused by the communication channel noise. The binary version of LDPC codes can distinguish the occurrence of bit flipping through information transmission by the channel. The theoretical limit of the channel error correction is known by a quantity called the channel capacity or the Shannon capacity of channel, which is the maximum possible rate under error free communication [4].

Nowadays, LDPC codes is being used in extensive areas such as the wireless communication systems [5], the cryptography based systems [6], the optical communication systems [7], the satellite communication systems [8], the digital video broadcasting systems [9], and so on. Furthermore, the LDPC codes are an important part of the Wi-Fi 802.11 standard technology as an optional part of 802.11n and 802.11ac [10]. These codes are efficient and faster than turbo codes because of their low decoding complexity [11].

The spread spectrum technique is a procedure that spreads the particular bandwidth of a communication signal carrying information to a wider bandwidth in the frequency domain [12]. There are various types of spreading schemes, including the direct sequence spreading, the frequency hopping spreading, the time hopping spreading, and also combination of them which cause hybrid schemes. These schemes are being used for different reasons and provide many advantages such as the secure communications, resistant to interference, the ranging ability, the low probability of interception, anti-jamming, noise protection, multiple-access communication, etc. [12]. In this work, we focus on the direct sequence technique. Clearly, our proposed scheme can be extended by using other techniques of spectrum spreading.

By employing the LDPC codes in the spread spectrum procedure, we combine their goals and obtain a joint scheme with their advantages. Specially, the results of BER performance get so much better by using this joint scheme. As a simple example, for a communication system with a parity-check matrix $H(128,256)$, using only LDPC coding procedure, the BER value at SNR=5 dB becomes about $3.1\times10^{-3}$, while by using both LDPC codes under the DSSS scheme with the processing gain of 10, the BER value at SNR=5 dB becomes about $1.3\times10^{-9}$ which improves the BER value significantly. In addition, by increasing the processing gain of spread spectrum, the BER gets better. Based on our experiments and implementations, if we consider a system with processing gain of 100 and the same LDPC code, then lower values of the BER are completely achievable in such a way that we observe the BER values close to zero. Furthermore, utilizing the matrix $H$ for generating spreading code of DSSS provides a low-complexity scheme which is another achievement.

The rest of the paper is organized as follows. The system model and definitions are proposed in Section II. Details of the LDPC codes, DSSS technique, and our proposed joint scheme are provided in Section III. The simulation and implantation results are in Section IV. Finally, Section V concludes this paper.

## II. SYSTEM MODEL AND DEFINITIONS

The task of a basic communication system is transmitting data from one point to another point. A block diagram of a communication system that is equipped with the spread spectrum technique has demonstrated in fig. 1.

A source and channel encoder is located prior to the modulator for preparing binary data. The source encoder block aims to compress data and remove redundancy. Conversely, the channel encoder adds some redundancy intentionally to protect compressed data against the communication channel noise which is considered to be an Additive White Gaussian Noise (AWGN) channel.

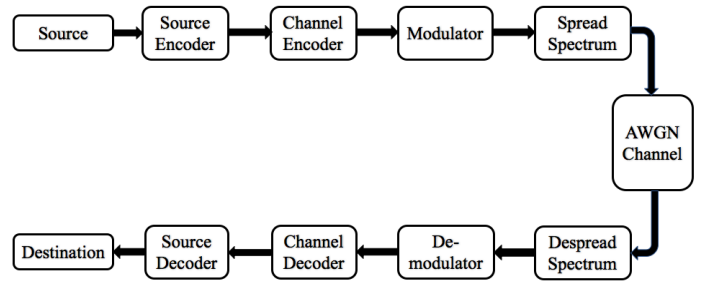

Fig. 1. A block diagram of a communication system with Spread-Spectrum procedure.

The source encoder maps the input information symbols to the code-words with a shorter length. It can be lossy or lossless. The Huffman codes, the Lemple-Ziv codes, the Fano codes, the arithmetic codes, and the Shannon codes are some of the applicable source coding schemes [13].

The channel encoder generates longer code-words that are made by the source encoder to increase the reliability of transmission over the channel. In another word, the channel coding schemes enable the receiver to detect and correct errors which occur during data transmission. The LDPC codes, the turbo codes, the family of convolutional codes, and the Reed-Solomon codes are some of the famous schemes that are used for the channel coding [14].

Data for being transferred on the channel needs to be superimposed on a high-frequency signal, which is called carrier wave, such that its frequency is higher than encoded data. Hence, the modulator in the communication system superimposes the encoded data on a high-frequency signal. By changing one of the carrier wave parameters such as amplitude, phase, or frequency, the scheme of modulation is determined. The phase-shift keying, the frequency-shift keying, and the amplitude-shift keying are methods of digital modulation [15].

The task of the spread spectrum block in the telecommunication system is spreading the signal over a large frequency band to provide secure efficient communication [12]. After spreading spectrum of the modulated data, it is transferred through the AWGN channel, which is an acceptable model for real-world communication channels.

At the receiver, all operations should be done reversely on the received signal based on processes, which have been done on the input data at the transmitter. Basically, all of the blocks in a communication systems act separately and independently, but they can be joined together for reaching more efficiency and lower complexity, e.g., joint source-channel encoder. In this paper, we jointly design channel encoder and spread spectrum.

## III. THE LDPC BASED DSSS SYSTEMS

The LDPC codes are linear block codes that use a low-density parity-check matrix $H$, such that the portion of ones to zeros is very low. The parity-check matrix $H$ has some properties which make it useable for a variety of intentions. By increasing size of parity-check matrix $H$, the performance improves due to decreasing the BER [5]. Furthermore, by increasing the number of iterations in the Sum-Product Algorithm (SPA) in the receiver the BER performance improves [5]. Also, by decreasing the rate of the parity-check matrix $H$, the BER improves [5]. An example of a parity-check matrix and its related Tanner graph is demonstrated in fig. 2, that its corresponding graphical representation of the parity-check matrix $H$. Tanner graph has two groups of nodes, the variable nodes and the check nodes, $j$-th variable node and $i$-th check node are connected each other if and only if $h_{i,j}=1$. The close path that is shown by bolded lines in fig. 2, shows a loop of length=6 in the parity-check matrix $H$. Girth of a parity-check matrix $H$ is the shortest loop which exist in its Tanner graph. Performance of LDPC codes improve by designing a parity-check matrix $H$ with a higher Girth [16].

$$H = \begin{pmatrix} 1 & 1 & 1 & 1 & 0 & 0 & 0 & 0 & 0 & 0 \\ 1 & 0 & 0 & 0 & 1 & 1 & 1 & 0 & 0 & 0 \\ 0 & 1 & 0 & 0 & 1 & 0 & 0 & 1 & 1 & 0 \\ 0 & 0 & 1 & 0 & 0 & 1 & 0 & 1 & 0 & 1 \\ 0 & 0 & 0 & 1 & 0 & 0 & 1 & 0 & 1 & 1 \end{pmatrix}$$

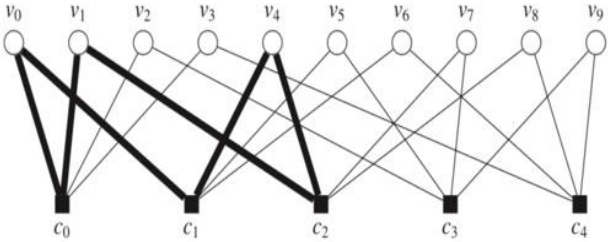

Fig. 2. Tanner graph of the parity-check matrix $H$.

Assume $C$ is the output of the channel encoder. At the transmitter, the output $C$ obtains by multiplying the binary information block of length $n$ and the generator matrix $G$ with the size of $n \times m$. The length of the output vector becomes $m$ such that, the result of multiplying the output vector $C$ and transpose of the parity-check matrix $H$ becomes an all-zero vector ($C.H^T=0$). Clearly, the length of the redundancy that is added to the codeword is $m-n$. Note that, the generator matrix $G$ is derived from the parity-check matrix $H$ [7]. Therefore, the output $C$ is as follows,

$$C_{1,m} = D_{1,n} G_{n,m} . \qquad (1)$$

where, $D_{1,n}$ is information bits. At the receiver, the SPA is being used for channel decoding because of its good performance and low-complexity. The SPA is a kind of bit-flipping algorithm, and the only difference is that the SPA is soft decision message-passing algorithms while bit-flipping algorithm is a hard decision algorithm [17]. Message-passing algorithms are iterative algorithms such that some messages pass between the variable nodes and the check nodes iteratively, until the process stops. Suppose $R_i$ as the received data and $E_{j,i}$ of probability matrix $E$ as the probability of being $R_i=1$.

The probability $P$ of being established a parity-check equation such that, $R_i$ equals 1 can be calculated as follows where $B_j$ is the set of variable nodes that has connected to $j$-th check node.

$$P_{j,i}^{ext} = \frac{1}{2} - \frac{1}{2} \prod_{i' \in B_j, i' \neq i}(1 - 2P_{j,i'}) . \qquad (2)$$

Log-Likelihood Ratio (LLR) is a metric for a binary variable and can be written as follows, the sign of LLR is a metric for making a hard decision and, the magnitude of LLR shows the reliability of the made decision,

$$L(c) = \log \frac{P(r=0)}{P(r=1)} . \qquad (3)$$

Therefore, the probability of being $R_i$ equals 1 or 0 can be written as follows,

$$p(r=1) = \frac{e^{-L(r)}}{1+e^{-L(r)}} , \qquad (4)$$

$$p(r=0) = \frac{1}{1+e^{-L(r)}} . \qquad (5)$$

Transferring information from $j$-th check node to $i$-th variable node is expressed as follows,

$$E_{j,i} = L(P_{j,i}^{ext}) = \log \frac{1 + \prod_{i' \in B_j, i' \neq i}(\frac{1-e^{-L(P_{j,i'})}}{1+e^{-L(P_{j,i'})}})}{1 - \prod_{i' \in B_j, i' \neq i}(\frac{1-e^{-L(P_{j,i'})}}{1+e^{-L(P_{j,i'})}})} , \qquad (6)$$

$$= 2\tanh^{-1} \prod_{i' \in B_j, i' \neq i} \tanh(L(P_{j,i'})/2) .$$

Also, by assuming following relations,

$$S_{i,i'} = \text{sign}(L(P_{i,i'})), \qquad (7)$$

$$M_{i,i'} = |L(P_{i,i'})| , \qquad (8)$$

$$L(P_{i,i'}) = S_{i,i'} M_{i,i'} , \qquad (9)$$

the equation 6 can be written as follows,

$$E_{j,i} = \prod_{i'} S_{i,i'} \cdot 2\tanh^{-1} \log^{-1} \sum_{i'} \log(\tanh(\frac{1}{2} M_{j,i'})). \qquad (10)$$

The total LLR of the $i$-th bit is as follows, where, $A_i$ is the set of connected check nodes to the $i$-th variable node,

$$L_i^{\text{total}} = L_i + \sum_{j \in A_i} E_{j,i}. \quad (11)$$

The decoder is initialized by setting variable node information $L(P_{j,i})$ such that $h_{i,j}=1$. Then, $L_i$ for the binary input and AWGN channel can be calculated as follows, where, $y_i$ is the $i$-th information bit that has been received,

$$L_i = L(c_i|y_i) = -4y_i/N_0. \quad (12)$$

After calculating total LLR of variable nodes, $\hat{R}_i$ is obtained as follows, where, $N$ is the length of codeword,

$$\hat{R}_i = \begin{cases} 1 & \text{if } L_i^{total} < 0 \\ 0 & \text{else} \end{cases}, \text{ for } i=1, 2, \ldots, N; \quad (13)$$

The precision of decoded data $\hat{R}$ obtains by multiplying $\hat{R}$ and $H^T$. If $\hat{R}H^T$ equals zero, then the decoded data is correct, else LLRs should be updated as follows,

$$L(P_{j,i'}) = \sum_{j' \in A_i, i' \neq i} E_{j',i} + L_i. \quad (14)$$

This operation continues until multiplying of $\hat{R}$ and $H^T$ becomes zero vector, or the number of iterations equals the limit that has been supposed.

Spread spectrum is a procedure that spreads the particular bandwidth to a wider bandwidth in the frequency domain. The proportion of the chip rate to the data bit rate is called processing gain that is always higher than one. It means the chip rate of spreading sequence is higher than data bit rate ($R_c \gg R_d$). Due to the processing gain, the received signal has low probability of interception and can be demodulated even in the case that the power of the signal is lower than the level of noise power [8].

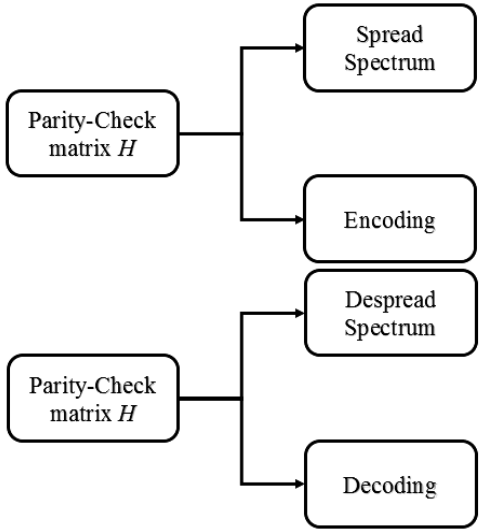

Fig. 3. The parity-check matrix $H$ for the proposed joint scheme.

In this paper, we generate the spreading sequence from the parity-check matrix $H$ by the exclusive disjunction of some parity-check matrix $H$ columns or by applying NOT operation on some parity-check matrix $H$ columns. By this procedure the channel encoder and the spread spectrum works jointly such that, it improves the performance of the communication system. Also, by using this procedure to generate the spreading sequence there is no need for transmitting the spreading sequence to the receiver because the receiver itself can generate the spreading sequence by the parity-check matrix $H$ that have. Then, there will be no need for blind estimation of spreading sequence that have shown in paper [9].

At the receiver, we generate the spreading sequence from the parity-check matrix $H$ exactly similar to the transmitter when the channel noise is applied. Then, we multiply the received data and spreading sequence. By multiplying the received data and spreading code, the de-spreading is done. Simultaneously, the spectrum of the channel noise spreads and its power decrease by the processing gain factor. Hence, the destructive effects of channel decrease and the performance improves.

## IV. SIMULATION AND IMPLEMENTATION RESULTS

In this section, we demonstrate and analysis the performance of using both LDPC codes and DSSS scheme jointly. The BER is calculated by dividing the number of incorrect received bits by the number of total received bits on the implementation tests.

All plotted charts in this section represents the BER performance of the LDPC encoding and decoding scheme with and without employing DSSS technique. The difference between two charts shows the improvement caused by the DSSS scheme. Figures 4, 5, and 6 represent the gap between two systems for different parity-check matrices of $H$. The number of iterations in the SPA is 100 and all reported values are based on average value over 50 tests. In our implementations, data modulation and spread spectrum modulation are both considered to be the BPSK modulation for its simplicity.

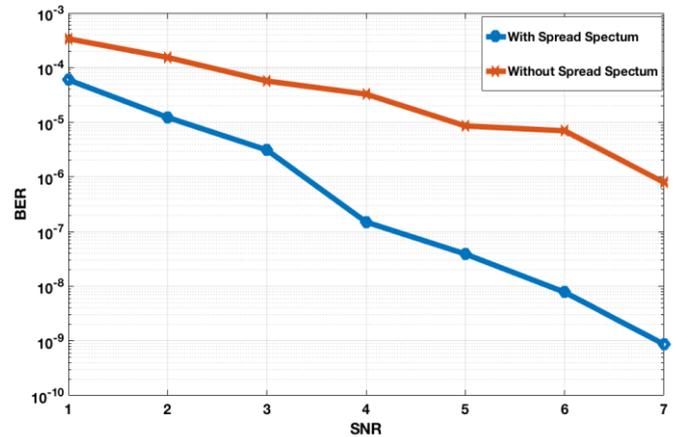

Fig. 4. Performance of the proposed scheme with parity-check matrix $H$(384,512,irregular), rate 0.25, and the processing gain is 4.

For instance, as it is observed from the figures, the BER of the system with the parity-check matrix $H(384,512)$ equipped with the spread spectrum improves. For instance, in a communication system with $H(384,512)$ by using only LDPC codes, the BER value at SNR=6 dB is about $7 \times 10^{-6}$, while it is equipped with the DSSS technique under the processing gain 4, the BER value at SNR=6 dB becomes about $7.8 \times 10^{-9}$.

Based on our results and observations, by increasing the SNR value, the amount of improvement in the BER value for the DSSS equipped system increases exponentially. The reason of this observation is that increasing the SNR cause to decrease the BER in the channel coding schemes, furthermore, the effect of noises weakens in the DSSS scheme, which cause another reduction in the BER value.

In the following, in figures 5 and 6, the performance of the proposed joint scheme are given for the parity-check matrices of $H(128,256)$ and $H(64,192)$, respectively. All degree distributions of the codes are designed based on the density evolution method [18], which produces irregular codes. Furthermore, the matrices are obtained using the progressive edge gross algorithm [18]. As it is seen, by increasing the SNR value, the amount of improvement in the BER value increases and this show the efficiency of our proposed scheme.

In fig. 7, we illustrate the performance of increasing the processing gain of the DSSS scheme. By increasing the processing gain of the DSSS the BER value decreases. For instance, the BER value in the proposed joint scheme with the processing gain of 10 is about $1.3 \times 10^{-9}$, while its value for the same communication system with the processing gain of 4 is about $8.2 \times 10^{-6}$. The exponentially increasing gap between the BER curves is clearly seen by increasing the SNR.

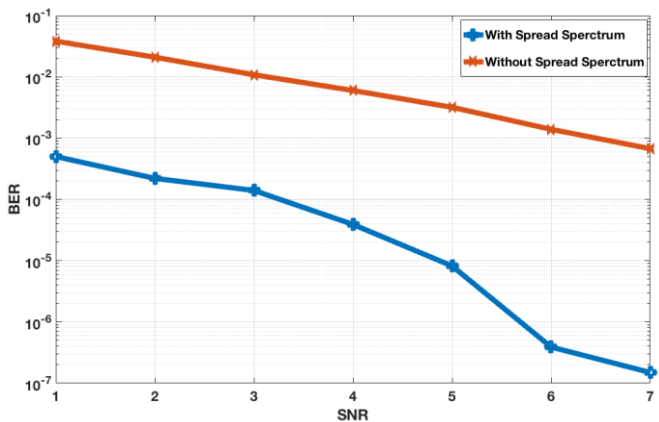

Fig. 5. Performance of the proposed scheme with parity-check matrix $H(128,256,\text{irregular})$, rate 0.5, and the processing gain is 4.

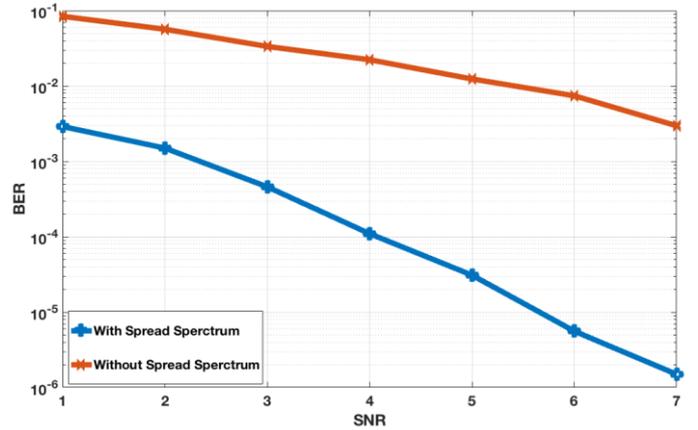

Fig. 6. Performance of the proposed scheme with parity-check matrix $H(64,192,\text{irregular})$, rate 2/3, and the processing gain is 4.

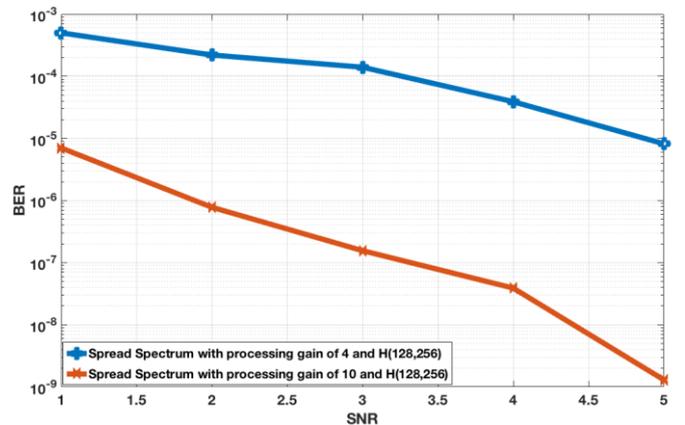

Fig. 7. The BER performance of the proposed scheme with the processing gains of 4 and 10 and parity-check code $H(128,256,\text{irregular})$, rate 0.5.

## V. CONCLUSION

An efficient joint channel coding and spectrum spreading scheme is proposed in this paper based on the LDPC codes and the DSSS technique. The parity-check matrix of an LDPC code which is employed in the encoding and decoding procedures is used to generate spreading code in a DSSS communication system. The BER performance of the system is significantly improved while the complexity does not increase a lot. For the high SNR values and also for high processing gain values of the spread spectrum systems, we observe very small values for the BER.